\documentclass[twocolumn,twoside,slac_two,nogroupedaddress]{revtex4}
\usepackage{graphicx}
\usepackage{fancyhdr}
\def\arcmin{\hbox{$^\prime$}}

\def\pflux {\mbox{ph cm$^{-2}$ s$^{-1}$}}

\begin{document}

\title{A variability study of the AGILE first catalog of $\gamma$-ray sources on 2.3 years of AGILE pointed observations }

%

\author{F. ~Verrecchia$^{1,2}$, C. Pittori$^{1,3}$, A.~Bulgarelli$^{4}$, A.W.~Chen$^{5}$, M.~Tavani$^{6}$, F.~Lucarelli$^{1,3}$, P.~Giommi$^{1}$, on behalf of the AGILE Collaboration}
\affiliation{$1$ ASI Science Data Center, ESRIN, I-00044 Frascati (RM), Italy}
\affiliation{$2$ Consorzio Interuniversitario Fisica Spaziale, villa Gualino - v.le S. Severo 63, I-10133 Torino, Italy;}
%
\affiliation{$3$ INAF-OAR, Astronomical Observatory of Rome, Monte Porzio Catone, Italy;}

\affiliation{$4$ INAF-IASF Bologna, via Gobetti 101, I-40129 Bologna, Italy;}
\affiliation{$5$ INAF-IASF Milano, via E. Bassini 15, I-20133 Milano, Italy;}
\affiliation{$6$ INAF-IASF Roma, via del Fosso del Cavaliere 100, I-00133 Roma, Italy;}

\begin{abstract}
AGILE pointed observations performed from July 9, 2007 to October 30, 2009 cover a very large time interval, with a $\gamma$--ray dataset useful to perform studies of medium to high brightness galactic sources in the 30 MeV -- 50 GeV energy range.
We present a study of the 1AGL galactic sources in the E\,$>$\,100 MeV band, over the complete Agile pointed Observation Blocks (OBs) archive.
 The first AGILE Gamma-Ray Imaging Detector (GRID) catalog included a sample of 47 sources (1AGL;~\cite{1AGL}), detected with a conservative analysis over the first year of operations dataset.
In the analysis here reported we used data obtained with an improved full Field of View (FOV) event filter, on a much larger (about 27.5 months) observation dataset, analyzing the merging of all data and each OB separately.
The data processing resulted in an improved source list as compared to the 1AGL one, particularly in complex regions of the galactic plane.
We present here some results on the revised 1AGL galactic sources and on the variability of some of them.

\end{abstract}

\maketitle

\thispagestyle{fancy}


\section{Introduction}

 AGILE (Astrorivelatore Gamma ad Immagini LEggero) (\cite{Tavani1}) is a mission of the Italian Space Agency dedicated to $\gamma$--ray astrophysics in the 30 MeV -- 50 GeV energy band, and simultaneously to hard X--ray in the 18 -- 60 KeV band. AGILE, in orbit since April 23 2007, has been the first instrument of a new generation of high-energy space missions based on the solid-state silicon technology, permitting to advance our knowledge on many source classes from active galactic nuclei, to pulsars (PSRs), unidentified $\gamma$--ray sources, Galactic compact objects and supernova remnants. On June 11, 2008 the Fermi Gamma-Ray Space Telescope (\cite{GLAST},\cite{LAT}) was launched, and is currently operating together with AGILE.

The AGILE Payload detector consists of the silicon tracker (ST; \cite{2002NIMPA.490..146B}, \cite{2003NIMPA.501..280P}), the X--ray detector SuperAGILE (\cite{2007NIMPA.581..728F}), the CsI(Tl) Mini-Calorimeter (MCAL; \cite{labanti09}) and an anti-coincidence system (ACS; \cite{2006NIMPA.556..228P}). The combination of ST, MCAL, and ACS forms the Gamma-Ray Imaging Detector (GRID). 
GRID is sensitive to photon energies in 30 MeV -- 50 GeV energy band, and thanks to silicon technology has a wide FOV (2.5 sr in pointing) and accurate timing (a few $\mu$s) and positional information (15$\arcmin$ location accuracy for $>$\,10$\sigma$ detection), with an angular resolution of 3.5$^\circ$ at 100 MeV and 1\,$^\circ$ above 1 GeV. 

\begin{figure*}[t]
\centering
\hbox{
\hskip -0.8truecm
\includegraphics[width=0.58\textwidth]{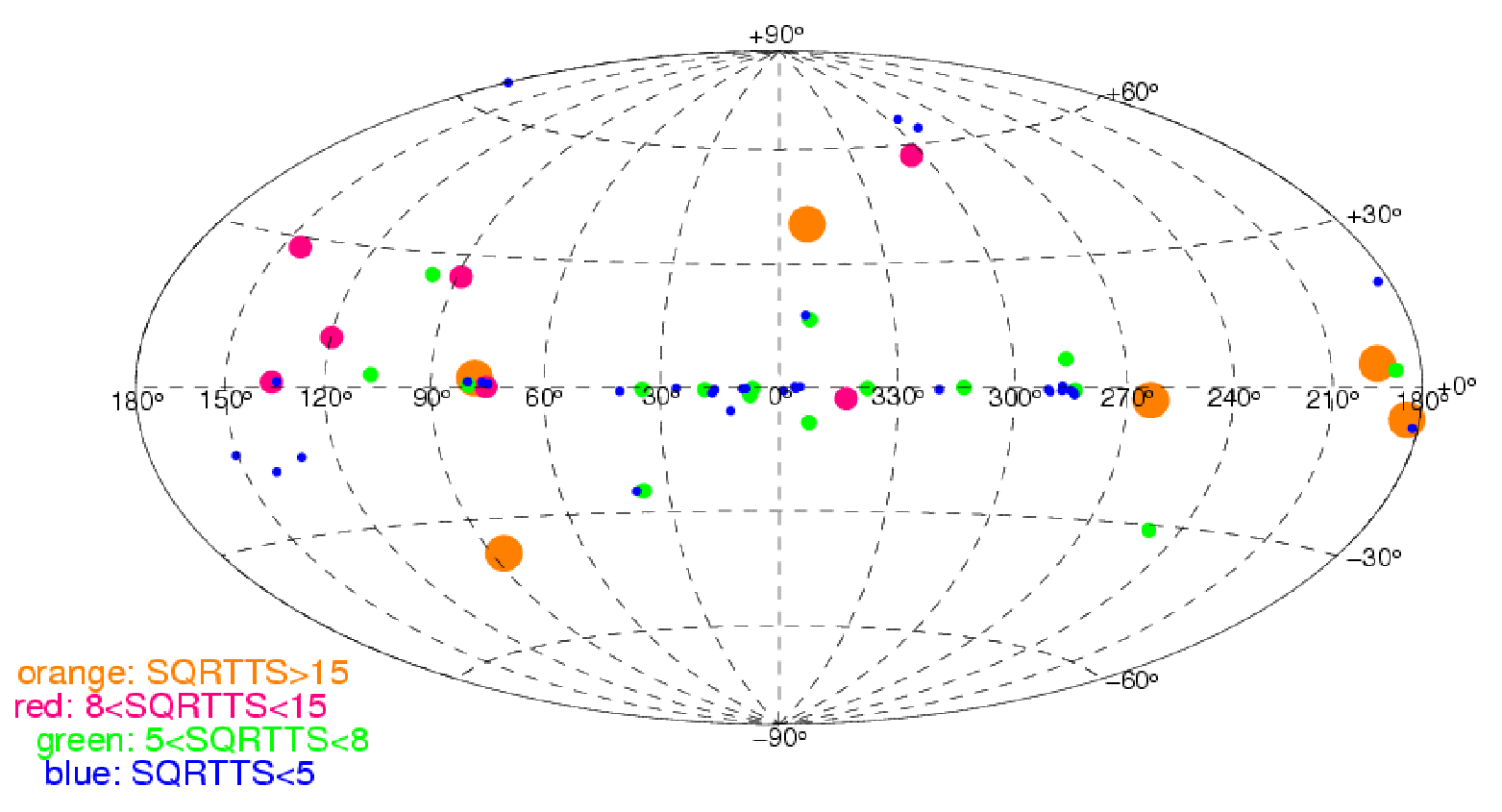}
\includegraphics[width=0.50\textwidth]{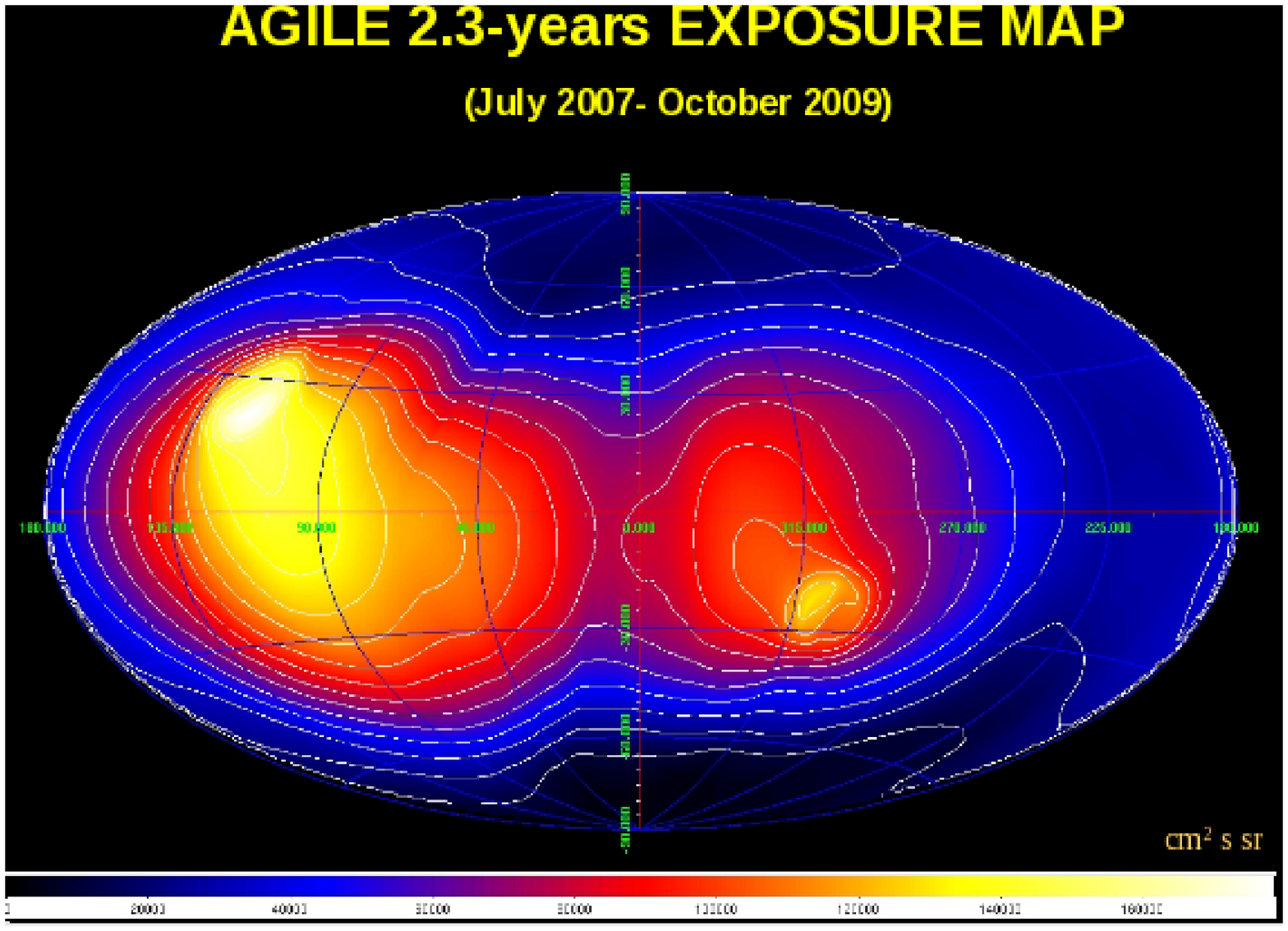}
}
\caption{ The Aitoff plot of the 62 distinct source positions detected in a revision of the original 47 1AGL ones, on all the pointed observations data (colors and symbol sizes are proportional to significance, blue the lowest and orange the highest). On the right the AGILE-GRID 2.3 years all-sky exposure map in Aitoff projection from all the pointed observations.} \label{f1}
\label{fig:Fig1}
\end{figure*}

 AGILE spacecraft operated in “fixed-pointing” mode since October 2009 (completing 101 pointings or “Observation Blocks”, OBs), when the attitude control system had to be reconfigured into ``spinning operation mode''. Currently the instrument pointing direction scans the sky with an angular velocity of about 0.8$^\circ$/s, accessing about 80\% of it each day. The AGILE satellite raw data, down-linked about every 100 minutes, are transmitted from Mission Control Center at Telespazio, Fucino, to the AGILE Data Center (ADC), part of the ASI Science Data Center (ASDC) located in Frascati (Italy). The ADC has the duties of data reduction, scientific processing and archiving and finally to distribute standard Level-2 data to Guest Observers (GOs) or, when data become public, to all the scientific community (see ADC web page http://agile.asdc.asi.it). \\

 The significance-limited (4 sigma) sample of 47 1AGL sources were detected in the E\,$>$\,100 MeV band with a conservative analysis of the inhomogeneous first-year sky coverage dataset. We present here the first results of a variability study of a sample of 62 sources (Fig. 1) analyzing separately each OB in the 2.3 years AGILE pointing mode dataset. The sample was obtained with a revision of the 1AGL sample on the maps obtained from the whole dataset (``deep'' maps from now on).

\section{OB source detection procedure}

  The standard analysis OB pipeline at the ADC was executed at the end of an OB to remove data corresponding to slews and occasional losses of fine-pointing attitude, and to build the official OB data archive. Moreover scientific maps in E$>$\,100 MeV energy band with a size of $60^\circ \times  60^\circ$, binned at $0.25 ^\circ$, were created selecting confirmed events and excluding albedo contaminated time intervals.
The ``pointing mode'' OB archive is composed of 101 OB covering the wide timespan with non uniform exposures (ranging from 1d to 45dd). All the OB were recently reprocessed with the last software release.
The procedure developed for this analysis on the whole archive in E$>$\,100 MeV band, is based on source detection at fixed preselected positions using the AGILE Maximum Likelihood (ML). 
It consists of two main steps: I) preliminary revision of the 1AGL source list, based on updated “deep” maps and on more recent published AGILE results; II) execution of a ML multi-source task on each OB data in a specific pipeline, keeping reference source positions fixed to those previously determined. In the analysis no automatic source detection process is implemented. 
A check leaving the position free is implemented to verify the goodness of candidate transient sources. Both steps were repeated for 5 main iterations to improve the multi-source detection in complex regions (see section \ref{sec:1rev}).

\subsection{Source detection method}

The detection method used is the ML. The significance (number of sigmas) of a source detection is given by the square root of the “test statistic” TS, defined as -2 times the log of the likelihood ratio, and expected to behave as $\chi$$^{2}$ (\cite{1996ApJ...461..396M}, Chen et al. in preparation). The likelihood ratio test is built considering for the background only hypothesis the AGILE diffuse $\gamma$--ray model (\cite{giuliani04}, Giuliani et al. in preparation). We used the task ``AG\_multi2'' included in the AGILE software for the ML ``multi-source'' analysis.

\subsection{The First AGILE catalog revision on the 2.3 yrs dataset}
\label{sec:1rev}

\begin{figure*}[ht]
\begin{center}
\hbox{

\includegraphics[width=0.35\textwidth]{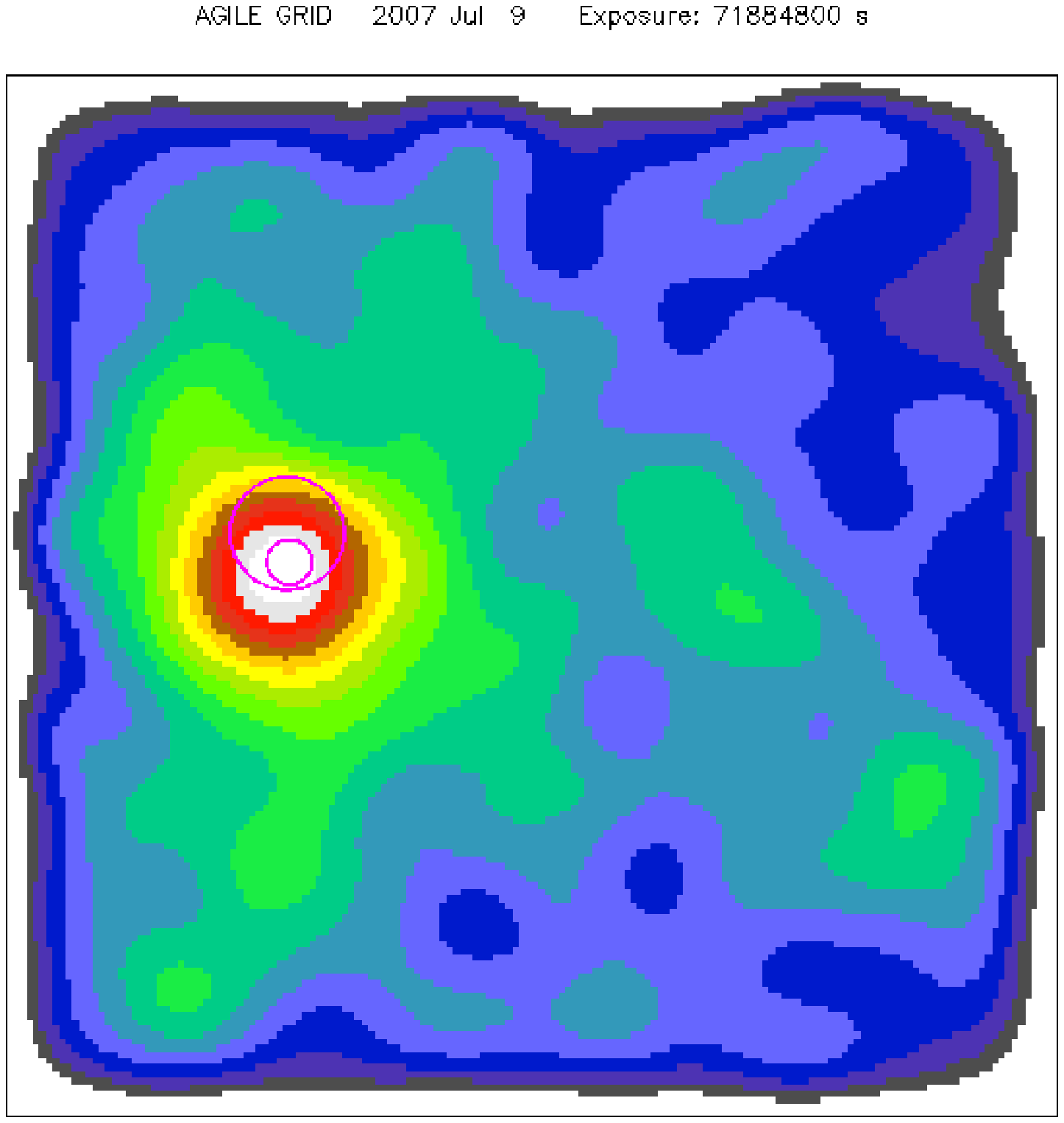}
\hskip +1.4truecm
\includegraphics[width=0.35\textwidth]{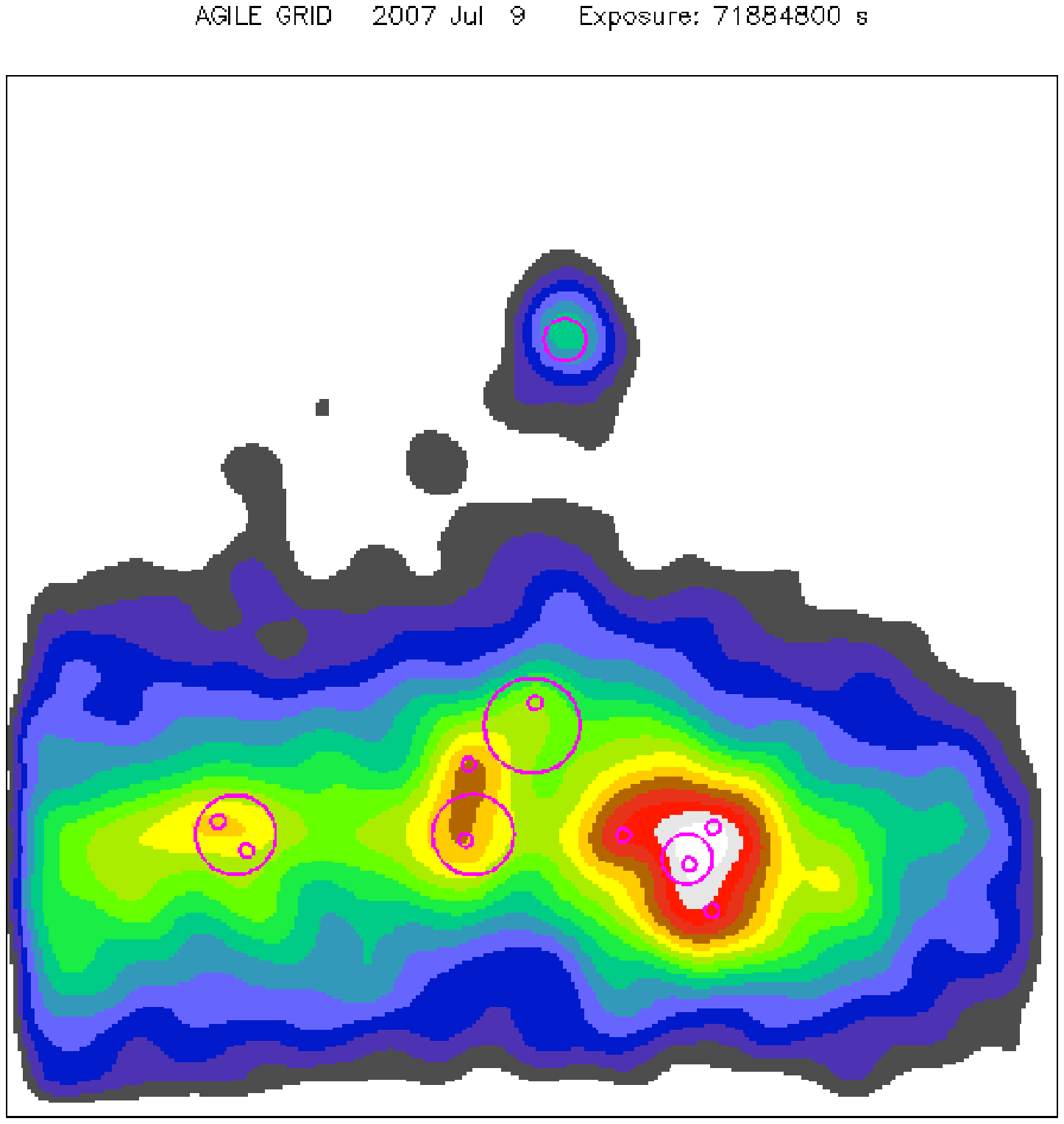}
}
\end{center}
\vskip -0.7truecm
\caption{ On the left a simple case of position refinement, the high latitude 1AGL J1846+6714. The new smaller error circle is indicated inside the 1AGL one. Right, a difficult position refinement case, the Carina region . New positions are indicated with small squares, with sizes not proportional to errors, while the 1AGL error circles are reported.
}
\label{fig:Fig2}
\end{figure*}

The first AGILE catalog (\cite{1AGL}) was built from data of the first year (July 2007\,--\, June 2008). The data analysis, based on a conservative event filter ([16]), and the non uniform sensitivity due to inhomogeneous first-year AGILE sky coverage, limited the results in complex galactic regions. A preliminary revision of 1AGL sources in regions such as the Carina, Cygnus, Crux and Galactic Center fields, was realized on updated merged maps from data up to October 2009 (see Fig. \ref{fig:Fig1}), using data from a new event filter (“FM3.119”; Bulgarelli et al in preparation). In this revision new “candidate” source were considered, and some with 3\,$<$\,$\sqrt{TS}\leq$\,4 were also included in the reference list to check their variability among all the OBs (see Fig. \ref{fig:Fig2}).

Moreover recent AGILE results obtained in these regions were taken into account (\cite{Tavanietacar}, \cite{TavaniCyg}, \cite{cygx1}, \cite{giulianiw28}, \cite{BulgCyg}). We considered also preliminary results from the ADC Quick Look processing on week timescale and the procedure results in the first 4 iterations. We obtained so an updated reference source list.

 A new source detection procedure on deep maps of all AGILE data in pointing mode has been realized for the next AGILE catalog and results will be reported in Bulgarelli et al. 2011.

\begin{figure}[ht]
\begin{center}
\hbox{
\hskip -0.8truecm

\includegraphics[angle=-90,width=0.52\textwidth]{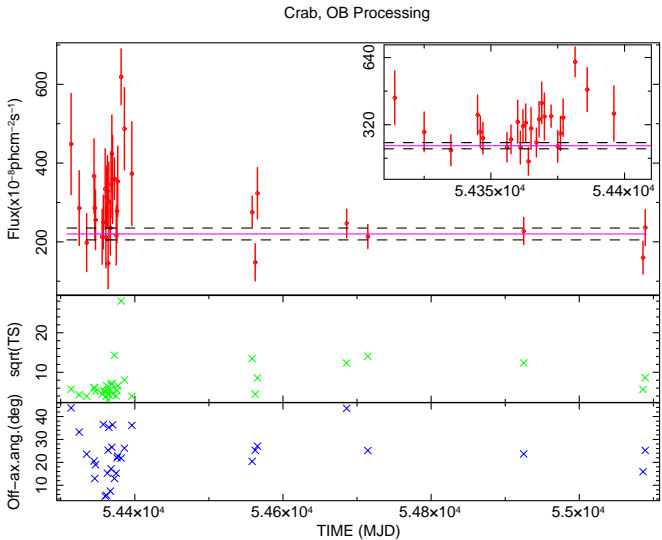}
}

\end{center}
\vskip -0.5truecm
\caption{
The Crab E$>$100 light curve (in 10$^{-8}$ \pflux; upper panel) and the plot of $\sqrt{TS}$ and the off-axis angle (in degrees) vs time (in MJD). The 1AGL flux value and the 1 sigma error levels are shown as magenta and black dashed lines. The 2007 flare episode is the most significant detection, at a flux of 619\,$\pm$\,71 \pflux.
}
\label{fig:Fig3}
\end{figure}



\section{Detection selection and variability analysis}

The results from the last execution (the fifth iteration) of the ML multi-source task on sources in the reference list, were filtered taking into account off-axis angle and significance. Our goal was to investigate the variability within the OB dataset for established sources on deep maps, and possibly search for flare episodes for some low significance source. As final selection only those sources having at least one detection at significance higher than 3$\sigma$ were accepted, for all sources having high significance ($>=$4) on deep maps, or otherwise with at least one detection above 4$\sigma$. Then all detections with $\sqrt(TS)\,\geq$\,2 were considered in the variability analysis. Source $\gamma$--ray flux variability among all OBs was tested according to the method developed by \cite{McL} and recently reported in an analysis of 1AGL J2022+4032 data (\cite{chen10}). The variability index V=-log(Q) is evaluated after the calculation of the weighted mean flux and its error and so also the relative $\chi$$^{2}$. Q is the probability of having a value of $\chi^{2}$\,$\geq$\,$\chi^{2}_{observed}$ for a source with constant flux. Sources with V$<$0.5 are usually classified as ``non variable'', with 0.5\,$\leq$V\,$<$1  as ``uncertain'' and with V$\geq$1 as ``variable''. Another index was calculated for comparison (used in Fermi catalogs; \cite{Abdo09}, \cite{Abdo10}), as a simple $\chi$$^{2}$. Both indices were computed also adding a systematic component of 10\% to flux errors  (V$_{sys}$, $\chi$$^{2}$$_{sys}$), to take into account the not well known particle background systematic, expected to vary among different integrations.

\section{Conclusions and future developments} \label{Res}

 Applying our selection criteria, we obtained a sample of 1267 detection of 62 distinct sources (Fig. \ref{fig:Fig1}; previous processing preliminary results were presented in \cite{VerC10}). For all sources a revision of single detections is on-going as a verification and improvement of the multi-source ML analysis and taking into account more recent calibrations.
 The non uniform exposures among the OBs, from 1 to 45 days, together with the pointing strategy, put strong constrains to the capacity of ``resolving'' complex regions in a large number of OBs and also to the variability analysis.

The light curve of the most bright sources with stable flux, such as Geminga and Vela PSR, have been preliminarily checked.  A particular case is the one of the Crab, which has been recently reported to have a ``non stable'' $\gamma$--ray emission (\cite{TavCrAAS}, \cite{TavCrab}, \cite{Abdo11}, \cite{VittCrab}). In Fig. \ref{fig:Fig3} is shown the Crab light curve showing the October 2007 flare episode, discussed in \cite{TavCrab}, at an higher flux compared to the 2010 flare. We obtained so a V$_{sys}$ of 0.4 for Geminga and 3.36 for Crab.

As an example of long-term monitoring light curves in the E$>$100 MeV band obtained in this analysis at the OB timescale, we show below (Fig.5 and 6) the results for the two known galactic sources 1AGL J1836+5923, associated to the LAT PSR J1836+5925, and 1AGL J0242+611 associated with the HMXB LS I+61303.

The use of V$_{sys}$ and $\chi$$^{2}_{sys}$ gave compatible results for most of the sources. The source class which included more variable sources were the bright Blazars according to both parameters, the most variable one being 3C 454.3.
The complete refined results on the OB timescale will be presented in Verrecchia et al. 2011.

\begin{figure}[ht]
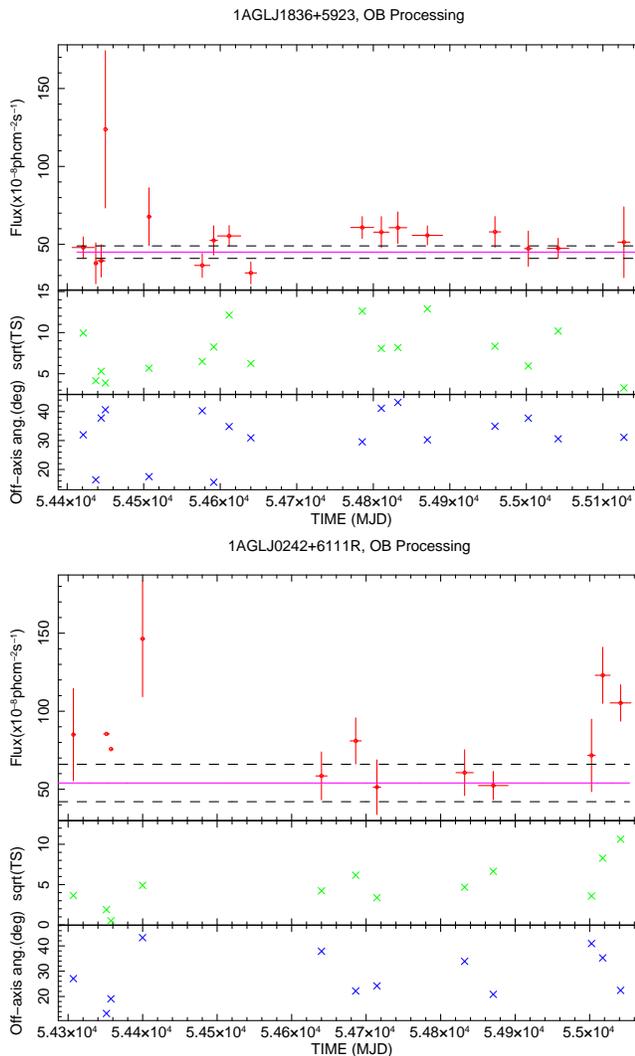

\begin{center}
\hskip -0.8truecm
\vbox{
\includegraphics[angle=-90,width=0.50\textwidth]{verrecchia7.ps}
\includegraphics[angle=-90,width=0.50\textwidth]{verrecchia8.ps}
}
\end{center}
\vskip -0.5truecm
\caption{ the 1AGL J1836+5923 and 1AGL J0242+611 light curves in E$>$100 MeV band, together with the $\sqrt{TS}$ and the off-axis angle plots. The magenta line is the 1AGL flux and the two black dashed lines are the levels for 1\,$\sigma$ errors. Sources have a V$_{sys}$ of 0.4 and 0.5 respectively.
}
\label{fig:Fig8}
\end{figure}




{\it Acknowledgements.} 
We acknowledge financial contribution from the agreement ASI-INAF I/009/10/0.
%

\bigskip 

\end{document}